\documentclass[12pt]{article}
\usepackage{graphicx}
\newif\ifagu\agufalse
\ifagu
\def\maketitle{}
\else
\def\authorrunninghead#1{}
\def\titlerunninghead#1{}
\fi
\def\gesim{\ \hbox to 0 pt{\raise .6ex\hbox{$>$}\hss}\lower.5ex\hbox{$\sim$}\ }
\def\lesim{\ \hbox to 0 pt{\raise .6ex\hbox{$<$}\hss}\lower.5ex\hbox{$\sim$}\ }
\newif\iftracking
\iftracking
\usepackage[normalem]{ulem}
\usepackage{color}
\long\def\versionone#1{\sout{#1} }
\definecolor{DullRed}{rgb}{.6,0.,0.}

\else
\long\def\versionone#1{}

\fi
\authorrunninghead{Hutchinson}
\titlerunninghead{Proton Distribution}
\begin{document}
\title{Near-lunar proton velocity distribution explained by electrostatic
  acceleration}

\author{I H Hutchinson\\
Plasma Science and Fusion Center and\\
Department of Nuclear Science and Engineering,\\
 Massachusetts Institute of Technology, Cambridge, MA 02139, USA}
\date{}
\maketitle

\begin{abstract}
  The observation of parallel ion velocity in the near-lunar wake
  approximately equal to external solar wind velocity \emph{can} be
  explained within uncertainties by an analytic electrostatic
  expansion model. The one-dimensional model frequently used is
  inadequate because it does not account for the moon's spherical
  shape. However, application of a more recent generalization to
  three-dimensions of the solution along characteristics predicts
  higher velocities, and is probably sufficient to account for the
  SARA observations on the Chandrayaan-1 space-craft.
\end{abstract}

\ifagu
\begin{article}
\fi

\section{Introduction}

Recent very detailed observations by the SARA experiment on board the
Chandrayaan-1 satellite have provided excellent documentation of
proton velocity distribution functions in the deep wake of the moon
when it lies in the solar wind, as reported by \cite{Futaana2010}. For
the conditions studied there, the interplanetary magnetic field is oriented
almost perpendicular to the solar wind and in the ecliptic. At an
altitude of 100km, above the moon's surface ($R_L \approx 1730$km),
and at an angle of 50 degrees to the terminator, on the equatorial
plane, the proton velocity distribution is observed to be concentrated
in a beam whose parallel velocity is approximately equal to its
perpendicular velocity. The perpendicular velocity is the solar wind
cross-field drift velocity approximately 325km/s in the downstream
direction. And the parallel velocity is inward along the magnetic
field.

The authors observe that this parallel velocity cannot be explained by
the one-dimensional self-similar expansion-into-vacuum solution
reviewed by \cite{Samir1983} (but developed substantially earlier, see
for example \cite{Gurevich1969}). That theory gives a velocity too small
by a factor of approximately 2, and density too large by a factor of
2-20. Alternative electron (kappa) distributions, which have been
similarly applied by \cite{Halekas2005}, do not resolve the
problem. This inconsistency has motivated further, more detailed,
modelling based on electromagnetic hybrid computation
\cite{Fatemi2012}, although without a clear resolution of the
discrepancies. The purpose of this brief communication is to point out
that the one-dimensional analytic model effectively treats the moon as
a planar disk rather than as a sphere. This approximation, while
reasonable for the distant wake, is not appropriate for these
near-wake situations. However, there exists an extension of the
analytic approach to 3-dimensions,
\cite{Hutchinson2008a,Hutchinson2008b} which provides results of
equivalent accuracy for objects of essentially arbitrary shape, and in
particular for a sphere. This theory, whose formulas are equally
compact, is outlined briefly and applied to the situation at hand,
predicting substantially higher parallel velocity. Probably, within
the approximation of neglecting the ion Larmor radius and other
uncertainties, it is sufficient to explain the observed results by
self-consistent electrostatic acceleration.

\section{Theoretical drift treatments}

The self-consistent one-dimensional quasi-neutral expansion of a
plasma into vacuum gives rise to self-similar solutions that are a function of
distance $s$ and time $t$. These can be applied to the wake of a
semi-infinite planar object past which plasma is flowing at a constant
drift velocity $v_\perp$, by identifying $s$ with distance parallel to
the magnetic field, and $t v_\perp$ with distance perpendicular to
magnetic field in the downstream drift direction. The self-similar solution is
a function only of the ratio $s/t$, and can be written:
\begin{equation}
  \label{eq:onedsoln}
  v_\parallel =  c_s + s/t .
\end{equation}

The general three-dimensional expression for arbitrary object shape
(of which the above is a special case) is more conveniently expressed
by normalizing all velocities to the sound speed $c_s\equiv
\sqrt{ZT_e/m_i}$ and writing the Mach numbers $M=v/c_s$. Then
\begin{equation}
  \label{eq:threed}
  M_\parallel = 1 + M_\perp \cot\theta .
\end{equation}
\begin{figure}[htp]
\hbox to \hsize{\hss
    \includegraphics[width=2.7in]{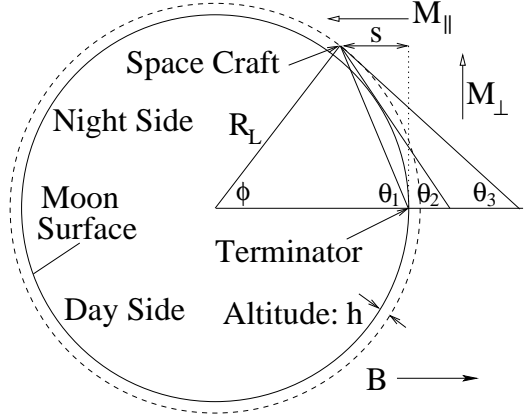}
\hss}
\caption{Geometry of the observations in the equatorial plane, which
  contains both the field (B) and the solar wind ($M_\perp$)
  directions.\label{geomfig}}
\end{figure}
Perpendicular ($\perp$) and parallel ($\parallel$) refer to components
in directions relative to the magnetic field. The ion motion
perpendicular to the field is presumed to be accounted for by drifts
arising from the self-consistent electric field, but is effectively
the external drift of the solar wind, which is what defines $M_\perp$,
a constant. The angle $\theta$, which is the controlling variable in
the solution, is shown rigorously by integration of the ion momentum and
continuity equations to be the angle to the magnetic field of one of
the ``characteristics'' of the differential equations. The
characteristic is that straight line passing through the location of
interest which is tangent to the surface of the object.

Figure \ref{geomfig} illustrates the definition of $\theta$. The
tangential characteristic's angle for the space-craft position is
$\theta_2$.  Ignoring finite gyro-radius, $\theta_2$ is the angle that
ought to be used in eq (\ref{eq:threed}). The angle that corresponds
to the one-dimensional approximation treating the moon as a disk, is
$\theta_1$, whose cotangent is $s/tc_sM_\perp$. [The tangent line to a
disk simply passes through its edge.] The angle $\theta_3=\pi/2-\phi$
is what would be required if the effective altitude of the observation
($h$) were taken to be zero. In that case the tangent point is at the
space-craft. The longitudinal angle of the space craft from the
terminator in the reported observations was $\phi=50$ degrees.

In all cases the corresponding theoretical density is 
\begin{equation}
  \label{eq:density}
  n = n_\infty \exp(M_{\parallel\infty} - M_\parallel)
\end{equation}
where $M_{\parallel\infty}$ is the component of external wind velocity
parallel to the field, equal to zero in this case.

The most notable new feature of the three-dimensional solution is that
the parallel velocity can become very large for positions near the
surface on the downstream side where $\theta$ may approach zero.

\section{Theory-Observation Comparison}

Simply evaluating the geometry gives
the following values of predicted $M_\parallel$ for the three angles
shown in Fig.\ \ref{geomfig}.
\begin{table}[htp]
  \centering
  \begin{tabular}{l|ccc}
   \hline
  Case:    & $\theta_1$& $\theta_2$& $\theta_3$ \\ \hline 
   $\theta^o$ & 68  & 59 &  40 \\
  $\cot\theta$ & 0.395 & 0.600 & 1.19 \\
  $M_\parallel/M_\perp$ & 0.495 & 0.700 & 1.29  \\\hline
      \end{tabular}
  \caption{Theoretical values for parallel velocity via eq
    (\ref{eq:threed}), for the three angle cases.}
\end{table}
\noindent
The final row is obtained by adding $1/M_\perp\approx 0.1$ to the
value of $\cot\theta$. This is based on the (cold-ion) thermal
acoustic speed in a proton plasma at the estimated electron
temperature (141,000K), approximately 35km/s, and the wind velocity
310-340km/s.  The uncertainty in the $1/M_\perp$ value is relatively
unimportant. Parallel velocity $M_\parallel/M_\perp\approx 1$ was
observed by the space craft, which is what the theory ought to match.

The one-dimensional approximation, $\theta_1$ gives velocity too low
by a factor of 2, as noted by \cite{Futaana2010}. However, using the
more appropriate three-dimensional formulation $\theta_2$, a
substantial increase of the predicted $M_\parallel/M_\perp$ to 0.7 is
obtained. This still does not fully reach the observed value. However,
the theory represented in these equations assumes that the gyro-radius
is negligible. This is not the case with these observations in which
the thermal ion gyro radius $\rho_i$ is approximately equal to the
altitude ($h=100$km). An over-correction of the finite gyro-radius
effect would be to consider the moon's effective collecting radius to
be not $R_L$ but $R_L+\rho_i$. That would give the $\theta_3$ case,
which over-predicts $M_\parallel/M_\perp=1.29$. Without detailed
three-dimensional kinetic calculation far beyond the scope of the
drift theory, it is not possible to estimate precisely the effect of
finite gyro-radius. It will increase the predicted velocity from the
$\theta_2$ case towards, but certainly no more than, the $\theta_3$
case. Incidentally, the drift analysis can be performed accounting for
the full parallel ion distribution function in finite ion temperature
cases (see \cite{Patacchini2009}), thus generalizing the iso-thermal
ion approximation used in the fluid theory. The predicted flow does not
change substantially.

The enhancement of $M_\parallel$ by a few, is more than sufficient
also to reduce the theoretical ion density, eq.\ (\ref{eq:density}), to
a level consistent with observations.

\section{Summary}

Analytic theory that accounts properly for the shape of the moon,
predicts substantially greater parallel ion velocity arising from
self-consistent electrostatic acceleration in the near-moon
wake. Within the uncertainty arising from finite ion thermal
gyro-radius (not included in the theory), and the other experimental
uncertainties, such a theory seems probably to be capable of
explaining the observed velocity. Undoubtedly there are many other
significant physical effects, including the dynamics of the magnetic
field that are represented in electromagnetic models. However, it
appears that the electrostatic structure of the wake is probably
sufficient to explain the parallel velocity observed by the
Chandrayaan-1 space craft.

\bibliography{Vacuum_Expansion,Lunar_Wake,Mine}

\ifagu
\end{article}
\fi

\end{document}